\documentclass[preprint2]{aastex63}
\pdfoutput=1 
\usepackage{amsmath,amstext}
\usepackage[T1]{fontenc}
\usepackage{apjfonts} 
\usepackage[figure,figure*]{hypcap}
\usepackage[colorinlistoftodos]{todonotes}
\usepackage{graphicx}
\usepackage[normalem]{ulem}

\shorttitle{LCBG Luminosity Function}
\shortauthors{Hunt et. al}

\begin{document}

\title{The Evolution of the Luminosity Function for Luminous Compact Blue Galaxies to z=1}

\author[0000-0001-8587-9285]{L.~R.~Hunt}
\affiliation{The United States Naval Observatory, 3450 Massachusetts Ave NW, Washington, D.C. 20392, United States, E-mail:lucas.hunt.ctr@navy.mil, lrhunt87@gmail.com}
\affiliation{Computational Physics Incorporated, 8001 Braddock Road, Suite 210, Springfield, VA 22151}
\author{D.~J.~Pisano}
\affiliation{Department of Physics \& Astronomy, West Virginia University, P.O. Box 6315, Morgantown, WV 26506, USA, E-mail: djpisano@mail.wvu.edu}
\affiliation{Center for Gravitational Waves and Cosmology, West Virginia University, Chestnut Ridge Research Building, Morgantown, WV 26505}
\affiliation{Adjunct Astronomer at Green Bank Observatory, Green Bank, WV, USA}
\author[0000-0002-8969-5229]{S.~M.~Crawford}
\affiliation{NASA Headquarters, 300 E Street SW, Washington, DC 20546, USA, E-mail:steven.m.crawford@nasa.gov}
\author[0000-0002-3131-4374]{M.~A.~Bershady}
\affiliation{Department of Astronomy, University of Wisconsin-Madison, 475 N. Charter Street, Madison, WI, 53706, USA, E-mail:mab@astro.wisc.edu}
\affiliation{South African Astronomical Observatory, PO Box 9, Observatory 7935, Cape Town, South Africa}
\affiliation{Department of Astronomy, University of Cape Town, Private Bag X3, Rondebosch 7701, South Africa}
\author[0000-0001-9058-4860]{G.~D. Wirth}
\affiliation{Ball Aerospace, 1600 Commerce St., Boulder, CO 80301, USA, E-mail:gregory.wirth@gmail.com}

\begin{abstract}
Luminous Compact Blue Galaxies (LCBGs) are compact, star-forming galaxies that are rarely observed in the local universe but abundant at z=1. This increase in LCBG number density over cosmic lookback time roughly follows the increase in the star formation rate density of the universe over the same period. We use publicly available data in the COSMOS field to study the evolution of the largest homogeneous sample of LCBGs to date by deriving their luminosity function in four redshift bins over the range $0.1\leq~z\leq1$. We find that over this redshift range, the characteristic luminosity (M$^{*}$) increases by $\sim$0.2 mag, and the number density increases by a factor of four. While LCBGs make up only about $18\%$ of galaxies more luminous than M$_{B}=-$18.5 at $z\sim0.2$, they constitute roughly $54\%$ at z$\sim$0.9. The strong evolution in number density indicates that LCBGs are an important population of galaxies to study in order to better understand the decrease in the star formation rate density of the universe since $z\sim1$. 
\end{abstract}

\section{Introduction}
\label{lcbg_lf_sec:intro}
The 4-meter telescope at the Kitt Peak National Observatory  identified a number of point sources whose optical emission lines did not match that of stars or QSOs \citep{Koo1988}. These compact objects were resolved in early Hubble Space Telescope observations using the Wide Field Planetary Camera \citep{Koo1994}, identifying compact, narrow emission line galaxies (CNELGs). Photometric and spectroscopic studies \citep{Phillips1997,Guzman1997} of these CNELGs in the flanking fields of the Hubble Deep Field found they evolve rapidly and contribute significantly to the star formation rate density of the universe near z=1. \citet{Werk2004} showed that the number density of these objects out to z=0.045 is roughly an order of magnitude lower than their number density at z=0.85 \citep{Phillips1997}, making them one of the most rapidly evolving galaxy types in the universe \citep{Rawat2007}. LCBGs' contribution to star formation rate density at z=1 \citep{Guzman1997}, combined with the correlation between their change in number density and the evolution in the star formation rate density of the universe from the local universe to z=1 means understanding of the nature and evolution of these galaxies provides important insight to galaxy evolution as a whole.

Compact narrow emission line galaxies \citep[CNELGs]{Koo1994,Koo1995,Guzman1996,Guzman1998}, faint compact galaxies \citep{Phillips1997,Guzman1997}, and blue nucleated galaxies \citep{Schade1995,Schade1996} are three classes of galaxies that have similar star formation rates, sizes and colors. While heterogeneous in terms of detailed morphology, these samples are similar in color and surface brightness, and for convenience were relabeled Luminous Compact Blue Galaxies (LCBGs). LCBGs are defined by having rest-frame M$_{B}<-18.5$ mag, B-V$<$0.6 mag , and SB$_{e}$(B)$<21$ mag arcsec$^{-2}$ \citep{Werk2004} so that  most galaxies in the original three samples are included, and future LCBG studies can target sources with the same selection criteria at all redshifts.

Other galaxy types that overlap with LCBGs in luminosity, color and surface brightness, have previously been described in the literature. \citet{Cardamone2009} showed that "Green Pea" galaxies, detected by the Galaxy Zoo project, overlap with LCBGs in blue luminosity, morphology, stellar mass, and metallicity. Higher mass LCBGs also overlap with Ultraviolet Luminous Galaxies (UVLGs), which were identified by \citet{Heckman2005} as being local counterparts to Lyman Break Galaxies. \citet{Guzman2003} and \citet{Hoyos2004} have both suggested that LCBGs could be low mass, lower redshift counterparts to the high redshift Lyman Break Galaxies. \citet{Phillips1997} and \citet{Guzman1997} noted that $\sim$40\% of their sample resemble local, vigorously star-forming HII galaxies. LCBGs also seem to be similar to Extreme Emission Line Galaxies \citep{Atek2014}, a class of low-mass star burst galaxies observed between 1.0<z<2.0 that contribute up to 34\% of the total star formation of emission line selected galaxies. Finally, we note that LCBGs are more massive \citep{Guzman1997,Garland2004,Tollerud2010}, luminous \citep{Garland2004}, and higher in metallicity \citep{Tollerud2010} than Blue Compact Dwarf galaxies (BCDs), and are therefore a distinct population.

Previous studies of LCBGs have concentrated on small snapshots in time, often being limited to low or intermediate redshift. Intermediate redshift (z$\sim$0.5) studies have looked at LCBG morphology \citep{Noeske2006} and spectral properties \citep{Tollerud2010}, environment \citep{Crawford2005,Crawford2014,Crawford2016,Randriamampandry2017}, and  number density \citep{Phillips1997,Guzman1997}. LCBGs make up about 50\% of the so-called Butcher-Oemler population of blue galaxies in clusters between redshift, 0.55<z<1 \citep{Crawford2005}, and are likely on their initial descent into a cluster \citep{Crawford2014}. Cluster and field populations of LCBGs show no significant difference in size, mass, luminosity, star formation rate, or metallicity \citep{Crawford2016}. \citet{Randriamampandry2017} found cluster LCBGs have a lower dynamical to stellar mass ratio than field LCBGs at intermediate redshift, indicating the mass of field LCBGs is more highly dominated by dark matter.  \citet{Guzman1997} and \citet{Phillips1997} found that LCBGs are similar to local HII star-forming galaxies, they constitute $\sim$45\% of the star formation rate density and $\sim$20\% of the galaxy population at z=0.6, and show no evolution in specific star formation rate. \citet{Tollerud2010} found LCBGs have a wide range of metallicities and show no correlation between size and stellar mass (M$_{*}$). 

Studies of local LCBGs have looked at number density \citep{Werk2004}, neutral gas content \citep{Guzman1997,Rabidoux2018}, kinematics from three-dimensional optical spectroscopy \citep{Perez-Gallego2011}, and morphology and environment \citep{Garland2015}. They have shown that LCBGs are locally rare \citep{Werk2004}, are rotationally supported \citep{Rabidoux2018, Perez-Gallego2011}, and their increase in star formation is not exclusively due to galaxy mergers \citep{Rabidoux2018,Garland2015, Perez-Gallego2011}. \citet{Garland2015} report that local LCBGs fit into roughly three categories: 20\% have star formation that is likely triggered by strong interactions, 40\% are clumpy spiral galaxies whose star formation is triggered by smoothly accreted gas from tidally disrupted companions or the intracluster medium, and 40\% are non-clumpy, non-spiral field galaxies with centrally concentrated morphologies, smaller effective radii, and smaller stellar masses. 

All previous studies of LCBGs show they are rapidly evolving, star-forming, compact galaxies. They are present in both dense and open environments out to at least a redshift of z=1. But previous studies of LCBGs have been limited in volume, concentrating either on small areas of the sky or small redshift ranges. LCBG studies using much larger samples provide robust statistics of the physical properties of these objects (e.g. average star formation rates, stellar masses, environment etc.) and can determine how comparable local LCBGs are to LCBGs at z=1.

We seek a comprehensive picture of the evolution of the number of LCBGs out to z=1 to supplement past and future studies. This study accomplishes that by using a large homogeneous sample of galaxies to create a catalog of LCBGs that is more than an order of magnitude larger than any previous LCBG catalog, and determines the evolution of their number density by deriving the luminosity function in four redshift bins between $0.1\le~z \le1.0$. In Section \ref{obs} we describe the data set we used, and how we selected LCBGs from it. In Section \ref{lcbg_lf_sec:LFE} we describe the 1/V$_{Max}$ method which we used to generate the luminosity function. In Section \ref{lcbg_lf_sec:COSMOS_LF} we discuss how our calculation of the luminosity function for the total galaxy population compares to previous work. In Section \ref{sec:Frac} we discuss the results from the LCBG luminosity function and compare them to other star-forming galaxy populations. Finally, in Section \ref{sec:Conc} we summarize our results.  Throughout this paper we adopt H$_{0}$=70 km s$^{-1}$ Mpc$^{-1}$, $\Omega_{M}=0.3$, $\Omega_{\Lambda}=0.7$ and we use the Vega magnitude system.

\section{Data}
\label{obs}
We use data obtained as part of the Cosmic Evolution Survey  \citep[COSMOS,][]{Scoville2008} with photometry rederived by the Galaxy and Mass Assembly (GAMA\footnote{GAMA is a joint European-Australasian project based around a spectroscopic campaign using the Anglo-Australian Telescope. The GAMA input catalogue is based on data taken from the Sloan Digital Sky Survey and the UKIRT Infrared Deep Sky Survey. Complementary imaging of the GAMA regions is being obtained by a number of independent survey programmes including GALEX MIS, VST KiDS, VISTA VIKING, WISE, Herschel-ATLAS, GMRT and ASKAP providing UV to radio coverage. GAMA is funded by the STFC (UK), the ARC (Australia), the AAO, and the participating institutions. The GAMA website is http://www.gama-survey.org/ .}) survey team. COSMOS is an HST Treasury Project \citep{Scoville2008} that used 590 orbits on the Hubble Space Telescope (HST) Advanced Camera for Surveys to survey a two square degree equatorial field. One of the key features of the COSMOS field is its accessibility by most astronomical facilities. This has allowed for surveys across the electromagnetic spectrum from radio \citep{Smolcic2017} to x-ray \citep{Civano2016}. The GAMA team are amassing large multi-wavelength photometric and spectroscopic datasets across numerous regions of the sky to study galaxy evolution. They have adopted the publicly available data from the COSMOS survey as their G10 region \citep{Davies2015a,Andrews2017}, the GAMA region at 10 hours right ascension.  

\subsection{Photometric data}
\label{sec:photdata}
The photometric dataset obtained by the COSMOS team covers the central 1 degree$^{2}$ of the 2 degree$^{2}$ COSMOS field, and includes observations from UV \citep{Zamojski2007a}, optical \citep{Capak2007,Taniguchi2007,Taniguchi2015}, and infrared \citep{McCracken2012a,Sanders2007,Oliver2012} bands. The GAMA team used their Lambda-Adaptive Multi-Band Deblending Algorithm in R \citep[LAMBDAR][]{Wright2016} to calculate consistent total flux photometry for all sources across all images to make this rederivation consistent with the other GAMA fields. We use the CFHT u$^{*}$ and Subaru B$_{j}$, V$_{J}$, r$^{+}$, i$^{+}$, z$^{+}$ photometry to calculate rest-frame magnitudes and colors, and the effective radii measurements were taken from HST F814W images \citep{Tasca2009} and matched to sources within 0.1". 

Our photometric sample includes galaxies with 15$\leq$i$^{+}\leq$22.5, roughly matching the photometric completeness limit for the original COSMOS observations \citep{Capak2007}. We adopt this photometric completeness limit in our study to avoid the need for a photometric incompleteness correction when deriving the luminosity function. The GAMA team developed a master star classification which adopts the COSMOS 2015 criteria for stellar classification \citep{Laigle2016}. This includes spectral energy distribution (SED) fitting, detection in infrared passbands, and proximity to the stellar sequence in BzK color space. They then apply a threshold in the size magnitude distribution to further discriminate between stars and galaxies. For more see section 3.5 of \citet{Andrews2017}. \citet{Laigle2016} has a classification for bright x-ray sources which \citet{Andrews2017} also suggests excluding from the sample as they likely contain some level of AGN activity which could lead to erroneous photometry for the galaxy. 

\subsection{Spectroscopic Data}
We used a compilation of spectroscopic redshifts gathered and reanalyzed by the GAMA team. The catalog uses redshift information from GAMA's AUTOZ \citep{Baldry2014}, zCOSMOS-bright 20k \citep{Lilly2009}, PRIsm MUlti-object Survey \citep{Coil2011,Cool2012}, VIMOS VLT Deep Survey \citep{Garilli2008}, Sloan Digital Sky Survey \citep{Ahn2014}, and the 30 band photometric redshift catalog \citep{Ilbert2009} to determine the best-fit redshift for each object. Full details of their reanalysis can be found in \citet{Davies2015a}. Objects were assigned quality flags of 1 to 4, 1 being robust, moderate resolution spectroscopic redshifts from VIMOS or SDSS, 2 being spectroscopic redshifts from the PRIMUS survey, 3 being uncertain spectroscopic redshifts, and 4 being photometric redshifts. We use the robust sample from \citet{Davies2015a}, which includes 16,130 objects with flags 1 \& 2 . We make corrections for incompleteness for 5,938 objects without spectroscopic redshifts which we discuss further in Section \ref{sec:Weighting}. 

\subsection{Rest-frame properties}
\label{sec:restframe}

To identify LCBGs, we had to determine their rest-frame properties including the absolute magnitude, surface brightness, and color of all galaxies in the GAMA G10 catalog. 

Absolute magnitudes were calculated using the equation:
\begin{equation}
M^{ref}=m^{obs}-DM(z,H_{0},\Omega_{m},\Omega_{\Lambda})-KC(z,SED)
\label{lcbg_lf_eq:AbsMag}
\end{equation}
In Equation \ref{lcbg_lf_eq:AbsMag} M$^{ref}$ is the reference band absolute magnitude, or the band in which we want to generate the luminosity function (in this case the Johnson B-band). m$^{obs}$ is the observed apparent magnitude of the object, selected to be closest to the rest frame B-band apparent magnitude for each object. DM is the distance modulus at redshift z defined as $5(\log_{10}(D_{L})-1)$, where D$_{L}$ is the luminosity distance for our adopted cosmology. KC is the k-correction defined as:
\begin{equation}
KC(z,SED)=[k^{ref}(z)+m^{obs}(z)-m^{ref}(z)]^{SED}
\label{lcbg_lf_eq:kcorrection}
\end{equation}
where k$^{ref}$(z) is the k-correction for the reference band assuming the object is at redshift z, m$^{obs}$(z) is the observed apparent magnitude, and m$^{ref}$(z) is the reference band apparent magnitude. The SED superscript indicates these values are taken from galaxy spectral template fits to observed galaxy fluxes using the standard \textit{kcorrect} package\footnote{http://kcorrect.org/} \citep{Blanton2006}. A detailed description of the spectral templates used and how they are used can be found in \citet{Blanton2006}. This method for calculating rest-frame absolute magnitude roughly follows the procedure from \citet{Ilbert2005}. 

\begin{figure}[htb]
\centering
\includegraphics[width=0.48\textwidth]{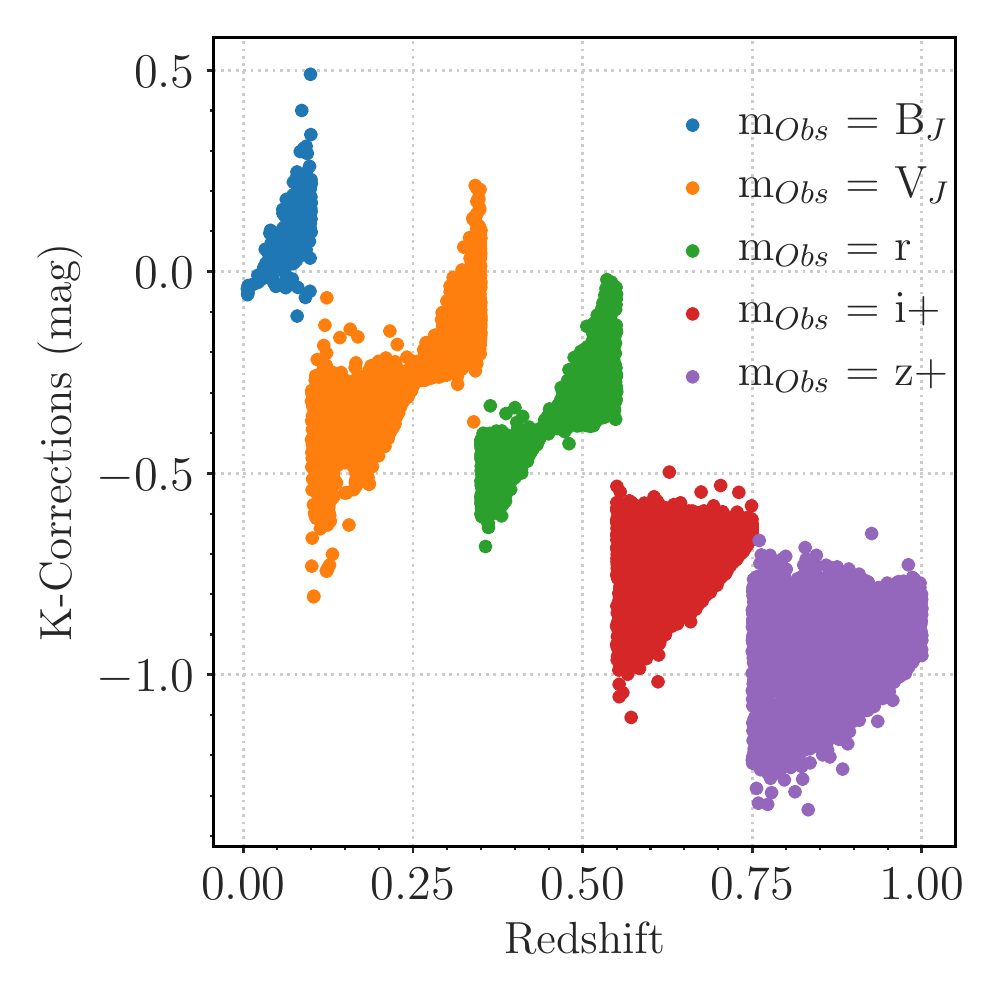}
\caption{k-corrections for all sources in our sample as a function of redshift.}
\label{lcbg_lf_fig:kc}
\end{figure}

We calculated the absolute magnitude using the observed filter closest to the rest-frame B band to minimize the size of  k-correction applied to the data. At higher redshifts,  the k-correction becomes more dependent on the spectral model with an increased chance of error due to mismatched models. Figure \ref{lcbg_lf_fig:kc} shows the range of k-corrections used at each redshift and how the distribution of the k-corrections narrows the closer the observed band is to the rest frame B-band

We also derived photometric offsets for the different filter responses between the G10 apparent magnitudes and the apparent magnitudes estimated by integrating the SED from the kcorrect code \citep{Blanton2006}.  Listed in Table \ref{lcbg_lf_tab:phot_off}, these corrections were applied to the observed apparent magnitudes.

We also use Equation \ref{lcbg_lf_eq:AbsMag} to calculate the rest-frame V-band absolute magnitude. We then used the rest-frame B-band and V-band absolute magnitudes to calculate the color ((B-V)$_{0}$) for each source.
\begin{table}[h]
\centering
\begin{tabular}{lc}
\hline
\hline
filter & offset\\
\hline
B$_{j}$........&$-$0.02\\
V$_{j}$........&0.05\\
r$^{+}$........&$-$0.04\\
i$^{+}$........&$-$0.01\\
z$^{+}$........&None\\
\hline
\hline
\end{tabular}
\caption{Photometric offsets between apparent magnitude in \citet{Andrews2017} and spectral templates from \citet{Blanton2006} \label{lcbg_lf_tab:phot_off} }
\end{table}

The effective surface brightness in mag arcsec$^{-2}$ was calculated using the equation below:
\begin{equation}
\mu_{e}=M_{B}+2.5\times \log_{10}(2\pi R_{e}^{2})+36.572
\label{lcbg_lf_eq:equation 3}
\end{equation}
where M$_{B}$ is the total absolute magnitude in the B band, and R$_{e}$ is the effective (or half-light) radius from \citet{Tasca2009} in units of kiloparsecs. This equation, taken from \citet{Graham2005}, was used to leverage our previous calculation of rest-frame absolute magnitudes, and takes into account the effects of surface brightness dimming.

\subsection{Sample Selection}
\label{sec:sampleselection}

As stated previously, \citet{Werk2004} define LCBGS as galaxies with M$_{B}\leq-$18.5 mag, $\mu_{e}(B)\leq$21 mag arcsec$^{-2}$, and (B-V)$_{0}\leq$0.6 mag. This combination of parameters provides the best distinction between intermediate redshift LCBGs and low luminosity irregular galaxies, elliptical galaxies, and normal spiral galaxies (See Section \ref{lcbg_lf_sec:intro} for more information). The selection criteria were based on source photometric properties to allow for selection at all redshift ranges and do not limit the sample in morphology or size.

\begin{figure}[htb]
\centering
\includegraphics[width=0.48\textwidth]{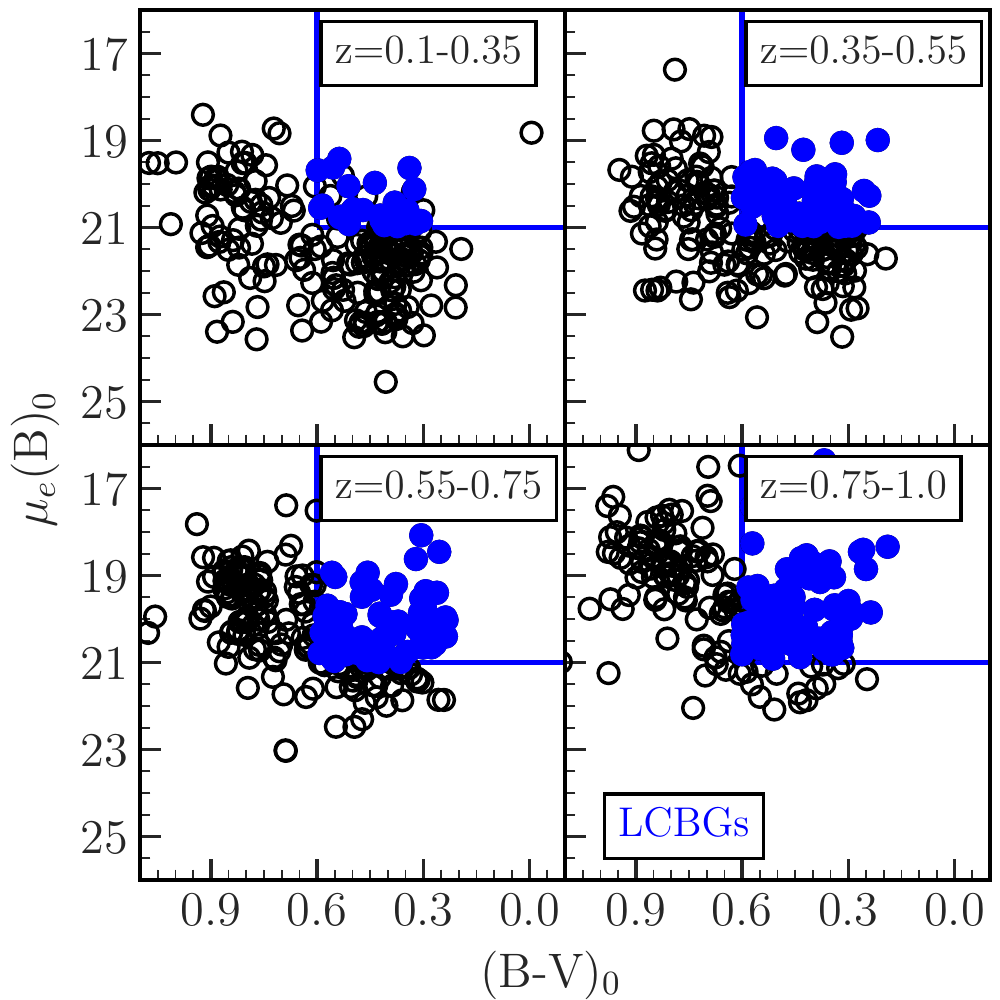}
\caption{Comparison of surface brightness and color selection criteria for a random subsample of 350 galaxies. All LCBGs (filled blue circle) will fall within the demarcated region of the B-band surface brightness vs color ($\mu_{e}$(B)$_{0}$ vs (B-V)$_{0}$) plot, but not all points in the demarcated region are LCBGs as some galaxies fail to meet the third LCBG selection criteria. All galaxies beyond z=0.55 will have M$_{B}$ <-18.5 due to the apparent magnitude limit of the survey and therefore are considered LCBGs if they meet the $\mu_{e}$(B)$_{0}$ and (B-V)$_{0}$ requirements.). All values are calculated in the galaxy's rest frame.}
\label{lcbg_lf_fig:sbc}
\end{figure}

\begin{figure}[htb]
\centering
\includegraphics[width=0.48\textwidth]{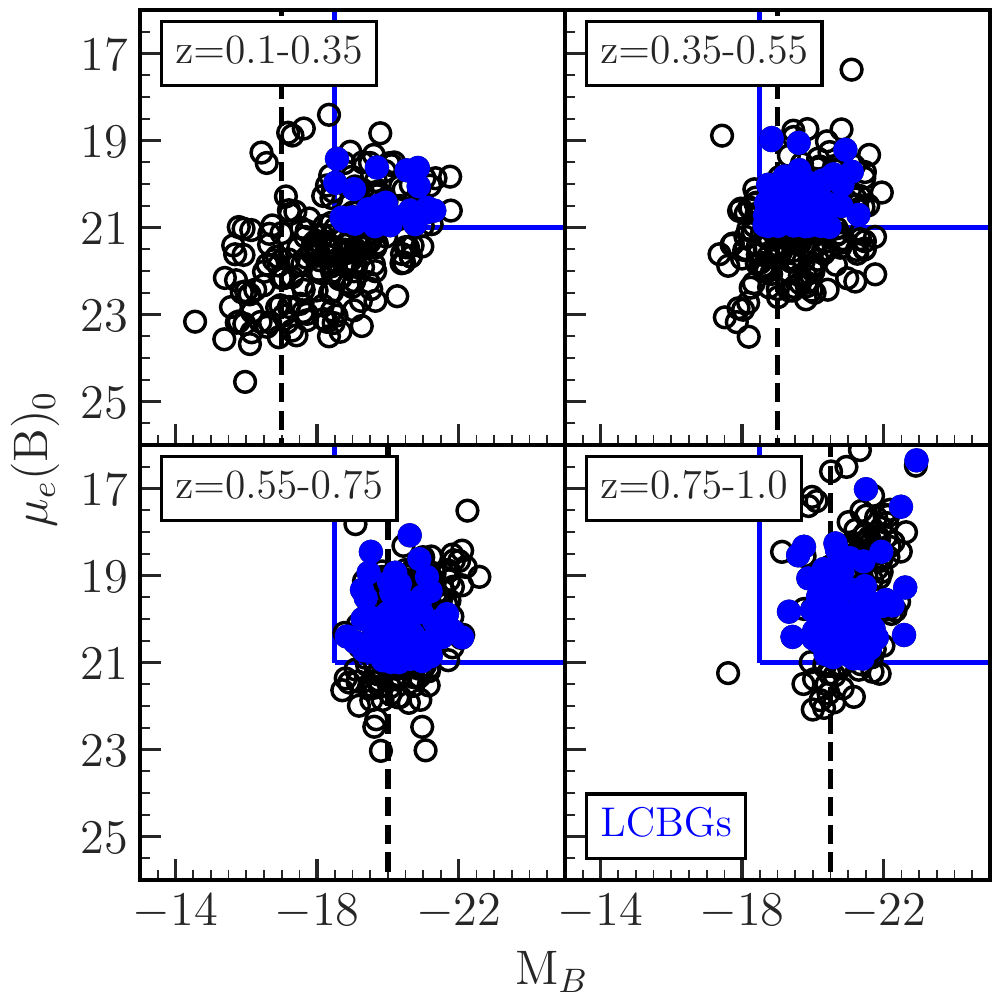}
\caption{Same as Figure \ref{lcbg_lf_fig:sbc} but for surface brightness vs absolute magnitude. The dashed line indicates the bias limit cutoff based on the apparent magnitude limits of the survey. The bias limit is discussed in more detail in Section \ref{lcbg_lf_sec:COSMOS_LF} }
\label{lcbg_lf_fig:msb}
\end{figure}

\begin{figure}[htb]
\centering
\includegraphics[width=0.48\textwidth]{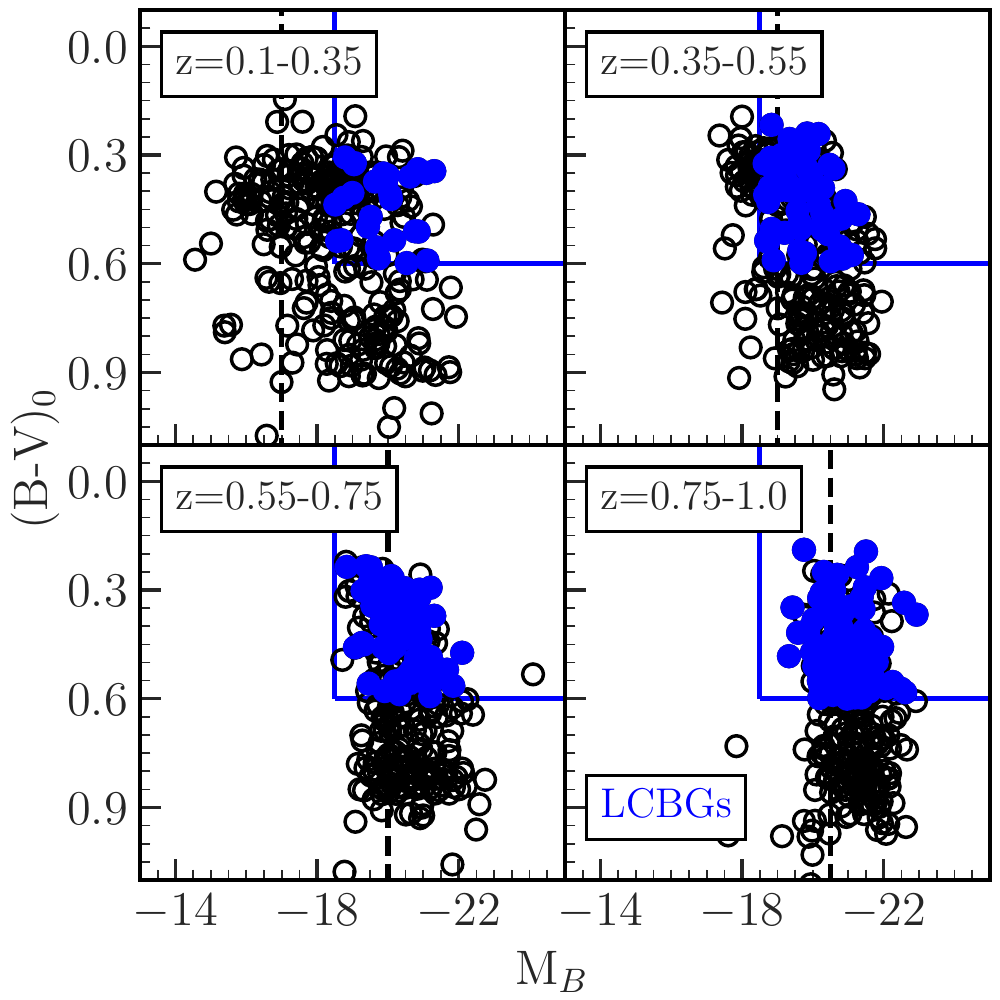}
\caption{Same as Figure \ref{lcbg_lf_fig:sbc} but for color vs absolute magnitude. The dashed line indicates the bias limit cutoff based on the apparent magnitude limits of the survey. The bias limit is discussed in more detail in Section \ref{lcbg_lf_sec:COSMOS_LF} }
\label{lcbg_lf_fig:mc}
\end{figure}

Figures \ref{lcbg_lf_fig:sbc},  \ref{lcbg_lf_fig:msb}, and \ref{lcbg_lf_fig:mc} show where LCBGs fall in $\mu_{e}(B)$-Color, $\mu_{e}(B)$-M$_{B}$ and Color-M$_{B}$ space respectively. \citet{Werk2004} point out that LCBGs are not distinct galaxies in  the luminosity-color-surface brightness parameter space, but exist at the extreme end of the continuum of galaxies in said space.  It is clear from these figures that not all galaxies in the demarcated regions are LCBGs because they do not meet the third selection criteria that is not displayed in the given plot.  The dashed line in Figures  \ref{lcbg_lf_fig:msb} and \ref{lcbg_lf_fig:mc} also shows the absolute magnitude bias limit cutoff \citep{Ilbert2004} discussed further in Section \ref{lcbg_lf_sec:COSMOS_LF}

\section{Luminosity Function Estimate}
\label{lcbg_lf_sec:LFE}
\subsection{Luminosity Function Estimator}
\label{lcbg_lf_sec:LFEstimator}
The luminosity function is defined as the number of galaxies per comoving volume, per magnitude bin, and is most often characterized using the Schecter  parameterization \citep{Schechter1976}. In magnitudes, the Schechter Function is defined as 

\begin{eqnarray}
&\phi(M)dM=&\\
&\frac{2}{5}~ln(10)~\phi^{*}~10^{0.4(M-M^{*})(\alpha+1)}\text{exp}(-10^{0.4(M^{*}-M)})dM\nonumber
\label{eq:equation 6}
\end{eqnarray}
where $\phi^{*}$ is the characteristic number density of galaxies per unit volume per unit magnitude, M$^{*}$ is the characteristic magnitude where the luminosity function transitions from an exponential into a power law, and $\alpha$ is the characteristic slope of that power law describing the faint end of the luminosity function.

Several methods have been developed to estimate the luminosity function \citep{Schmidt1968,Lynden-Bell1971,Turner1979,Choloniewski1986,Efstathiou1988} including the C$^{+}$ method \citep{Zucca1997}, a non-parametric, cumulative estimator, and the Sandage-Tammann and Yahil (STY) method \citep{Tammann1979}, a maximum likelihood estimator. For our analysis we used the 1/V$_{Max}$ method \citep{Schmidt1968} which counts galaxies within a known volume. We followed the method described by \citet{Willmer2006} and \citet{Ilbert2005}, which is briefly summarized below. The integral luminosity function for a given absolute magnitude is defined as: 
\begin{equation}
\int_{M_{bright}}^{M_{faint}}\phi(M)dM=\sum_{i=1}^{N_{G}}\frac{\chi_{i}}{V_{max}(i)}
\label{eq:equation 7}
\end{equation}
where $\chi_{i}$ is the weighting applied to correct for incompleteness (discussed in Section \ref{sec:Weighting}), N$_{G}$ is the number of galaxies in the absolute magnitude range M$_{faint}-$M$_{bright}$ and V$_{max}(i)$ is the maximum comoving volume within which a galaxy \textit{i} with absolute magnitude $M_{i}$ can be detected in the survey, defined as:
\begin{equation}
V_{max}(i)=\int_{\Omega}\int_{z_{min,i}}^{z_{max,i}}\frac{dV}{dz d\Omega}dz d\Omega
\label{eq:equation 8}
\end{equation}
Here again, z is redshift, and $\Omega$ is the solid angle of the survey region. In a magnitude limited survey, z$_{min,i}$ and z$_{max,i}$ are defined by:
\begin{equation}
z_{min,i}=\text{min}(z_{min},z(M_{i},m_{l}))
\label{eq:equation 9}
\end{equation}
\begin{equation}
z_{max,i}=\text{max}(z_{max},z(M_{i},m_{u}))
\label{eq:equation 10}
\end{equation}
where z$_{min}$ and z$_{max}$ are the lower and upper limits of the redshift bin the object occupies, and $z(M_{i},m_{l})$ and $z(M_{i},m_{u})$ are the redshifts at which an object with absolute magnitude M$_{i}$ would no longer fall in the apparent magnitude limits of the survey. We adopt Poissonian errors to describe the uncertainty in each of our luminosity function bins using: 
\begin{equation}
\sigma_{\phi}=\sqrt{\frac{\chi_{i}}{V_{max,}^{2}(i)}}
\label{eq:equation 11}
\end{equation}

\subsection{Weighting}\label{sec:Weighting}
\begin{figure}[htb]
\centering
\includegraphics[width=0.48\textwidth]{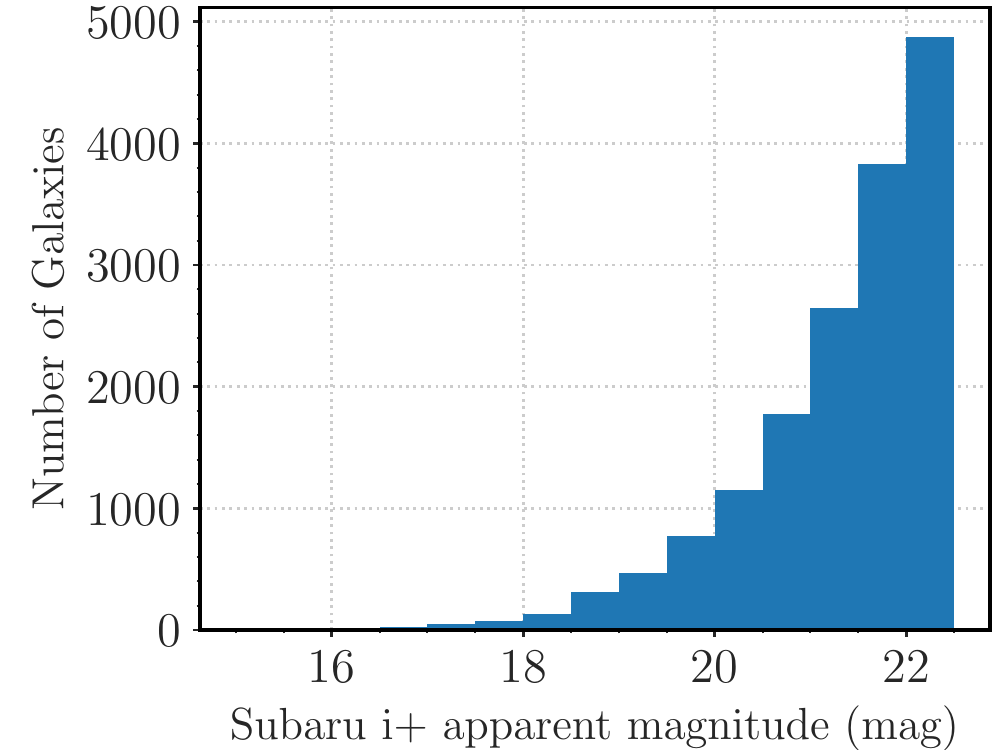}
\caption{Plot showing the distribution of apparent magnitude in the sample. The weighting was calculated in the bins shown in the Histogram. Since most of the higher weights are applied to the brighter objects, the high weighting only effects a small number of objects. }
\label{lcbg_lf_fig:app_mag_dist}
\end{figure}

We must correct for galaxies in the survey volume that have not been directly counted due either to photometric incompleteness (missing faint objects) or spectroscopic incompleteness (objects without spectroscopic redshifts). These incompleteness corrections are applied to all galaxies, including LCBGs, and all incompleteness corrections are calculated in Subaru i$^{+}$ band apparent magnitude bins of width 0.5 mag. The same weight is applied to all galaxies in an apparent magnitude bin and we have included Figure \ref{lcbg_lf_fig:app_mag_dist} to show how many sources are in each apparent magnitude bin. 

We include additional incompleteness corrections for objects in the survey for which we cannot calculate the surface brightness (objects that do not have a radius measurement) or color (objects that do not have a B or V band magnitude listed either due to the object not being visible in a given band or failure in the software) which only effect LCBG selection and therefore only apply to LCBGs. We adopted a similar photometric selection criteria to zCOSMOS, namely 15 mag$\leq$i$^{+}\leq$22.5 mag to maximize the number of sources with spectroscopic redshifts. This apparent magnitude range is also brighter than the photometric completeness limit of the G10 catalog, m$_{i^{+}}$=24.5 mag \citep{Andrews2017}, and we therefore do not adopt a photometric incompleteness correction.  

We made two corrections to compensate for spectroscopic incompleteness. First, we corrected for the photometrically identified galaxies that were not targeted in a spectroscopic survey. This weight is labeled as the target sample rate, and defined as:
\begin{equation}
w_{TSR}=\frac{N^{gal}_{Phot}}{N^{gal}_{Spec}}
\label{eq:equation 12}
\end{equation}
where $N^{gal}_{Phot}$ is the number of objects observed in the photometric catalog, and $N^{gal}_{Spec}$ is the number of objects targeted in the spectroscopic survey. As mentioned above, the spectra from this survey came primarily from the zCOSMOS survey \citep{Lilly2009}, and the PRIMUS survey \citep{Coil2011}. The zCOSMOS and PRIMUS surveys randomly targeted galaxies with m$_{i}$<22.5 mag and since the objects were randomly targeted, we do not make any corrections to the target sampling rate based on apparent magnitude or color.

We also corrected for objects that were targeted in the spectroscopic survey, but whose redshifts were unable to be definitively determined. This spectroscopic sampling rate weighting is defined as:
\begin{equation}
w_{SSR}=\frac{N^{gal}_{spec}}{N^{gal}_{spec}-N^{fail}_{spec}}
\label{eq:equation 13}
\end{equation}
where $N^{gal}_{spec}$ is the number of galaxies observed spectroscopically, and $N^{fail}_{spec}$ is the number of galaxies where the redshift was indeterminable. If we define the number of galaxies with secure redshifts as $N^{sec}_{spec}=N^{gal}_{spec}-N^{fail}_{spec}$, then the total weight of each object, $\chi_{i}$, is the multiple of the two weights described above:
\begin{equation}
\chi_{i}=\frac{N^{gal}_{Phot}}{N^{gal}_{spec}}\frac{N^{gal}_{spec}}{N^{gal}_{spec}-N^{fail}_{spec}}=\frac{N^{gal}_{Phot}}{N^{sec}_{spec}}
\label{eq:equation 14}
\end{equation}

The likelihood of obtaining a secure redshift varies with apparent magnitude as mentioned in \citet{Lilly2009}, so we calculate weights in Subaru i$^{+}$-band magnitude bins of 0.5. Weights generally range from 1.04 to 1.6. Six very bright objects have corrections above 5.5 due to small spectroscopic completeness of the brightest galaxies.   Only 5 of the 34 brightest galaxies had secure redshifts. Only 11 of the 34 brightest galaxies were targeted in a spectroscopic survey, and of those, 6 of the 11 targeted sources did not have secure redshifts. The low rate of spectroscopic coverage of the brightest sources may be due confusion or crowding of the brightest sources. 

We use a similar method to determine the extra weights for LCBGs. As mentioned in Section \ref{sec:photdata} in order to determine surface brightness to select for LCBGs we matched the photometry catalog from \citet{Andrews2017} to the morphology catalog from \citet{Tasca2009}. Since it's possible not all sources match between the two catalogs, the weight for surface brightness is:

\begin{equation}
w_{SBe}=\frac{N^{gal}_{phot}}{N^{gal}_{phot}-N^{gal}_{failrad}}
\label{eq:SBWeight}
\end{equation}
where N$^{gal}_{phot}$ is the number of galaxies identified in the photometric catalog  N$^{gal}_{failrad}$ is the number of galaxies that did not match between the photometric and morphological catalog.These corrections typically varied between 1.3 and 1.8 with only three LCBGs falling in an apparent magnitude bin that has a correction greater than 2. 

Finally, the color correction was applied when objects did not have a listed B and V-band apparent magnitude. The weight for color is: 

\begin{equation}
w_{B-V}=\frac{N^{gal}_{phot}}{N^{gal}_{phot}-N^{gal}_{failphot}}
\label{eq:COLORWeight}
\end{equation}
where N$^{gal}_{failphot}$ did not have B and V-band apparent magnitudes. The largest correction for color is 1.01 and most galaxies LCBGs did not have any correction for color. The total weight for LCBGs then becomes 
\begin{equation}
\chi_{i,LCBG}=w_{SSR}\times~w_{TSR}\times~w_{SBe}\times~w_{B-V}
\label{eq:equation 15}
\end{equation}

\begin{figure*}
\centering
\includegraphics[width=0.8\textwidth]{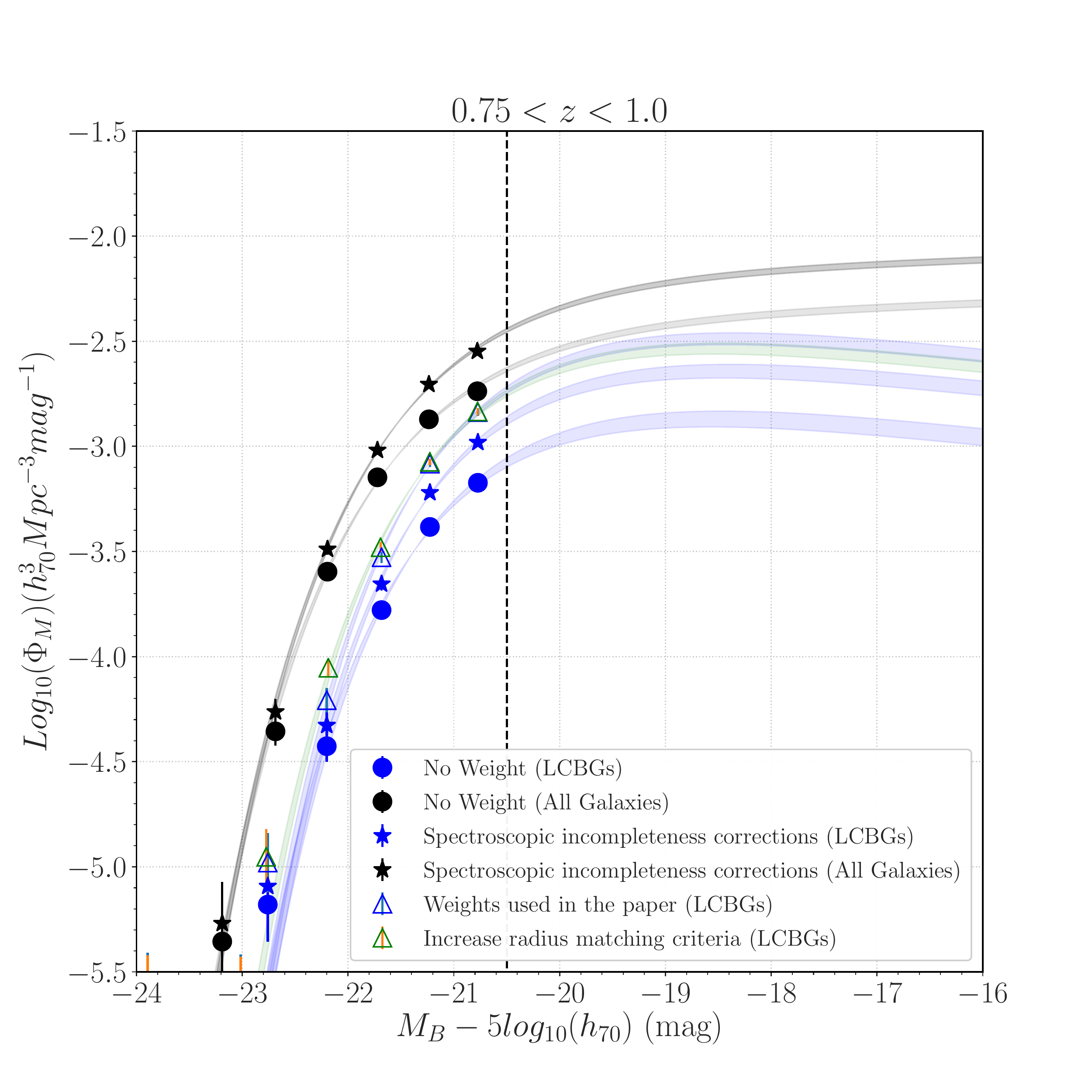}
\caption{This figure shows how the different weights added from Section \ref{sec:Weighting} effect the luminosity function for all galaxies (black symbols) and LCBGs (blue and green symbols). The luminosity function with no corrections for incompleteness is represented by circles. Corrections for spectroscopic incompleteness are represented by the stars. Surface brightness and color incompleteness corrections were included with the spectroscopic incompleteness corrections only to LCBGs and are represented by the blue triangles. Finally we relaxed the effective radius matching criteria to reduce the larger weighting for surface brightness incompleteness and that calculation is represented by the green triangles. The dashed line represents the bias limit.}
\label{lcbg_lf_fig:weight_compare}
\end{figure*}

In Figure \ref{lcbg_lf_fig:weight_compare} we show the affects of our weighting described in Section \ref{sec:Weighting} on the luminosity function in the highest redshift bin. We started by calculating the luminosity function for both the total galaxy population and LCBGs using no weighting in our incompleteness correction. Then we added weights to correct for spectroscopic incompleteness for the total galaxy population and the LCBG population. These were the only weights added to the total galaxy population, and were the final weights when calculating the total galaxy luminosity function. Finally we generated the luminosity function for LCBGs, adding weights to correct for galaxies without a calculated B-V color and galaxies for which we were not able to calculate the effective surface brightness. After applying all of the weights in the paper the number density of LCBGs increase by 129\% in the highest redshift bin Comparisons to previous measurements of the number density will be made in later sections. 

We also calculated the luminosity function for LCBGs after relaxing the matching criteria for the effective radius. As mentioned in \ref{sec:photdata} the effective radii measurements were matched to sources within 0.1". If we relax the matching criteria to sources within 0.3" we find more sources match, though we do get some cases where the same object in the morphology catalog is matched to multiple sources in the photometric/spectroscopic catalog. We then calculated the luminosity function with this relaxed matching criteria and plotted it in Figure \ref{lcbg_lf_fig:weight_compare}. The figure shows that the luminosity function of LCBGs with the applied weighting resembles the luminosity function generated with this relaxed matching criteria. The number density calculated using our weighting is 6.4\% higher than the number density calculated using the relaxed radius matching criteria in the highest redshift bin. The 68\% confidence intervals for the number density calculated using each method do overlap, and as such we believe that our derived luminosity function is robust to the details of how we match sources and our weighting scheme.

\section{Comparison to Previous Work}
\label{lcbg_lf_sec:COSMOS_LF}
\begin{figure}[htb]
\centering
\includegraphics[width=0.48\textwidth]{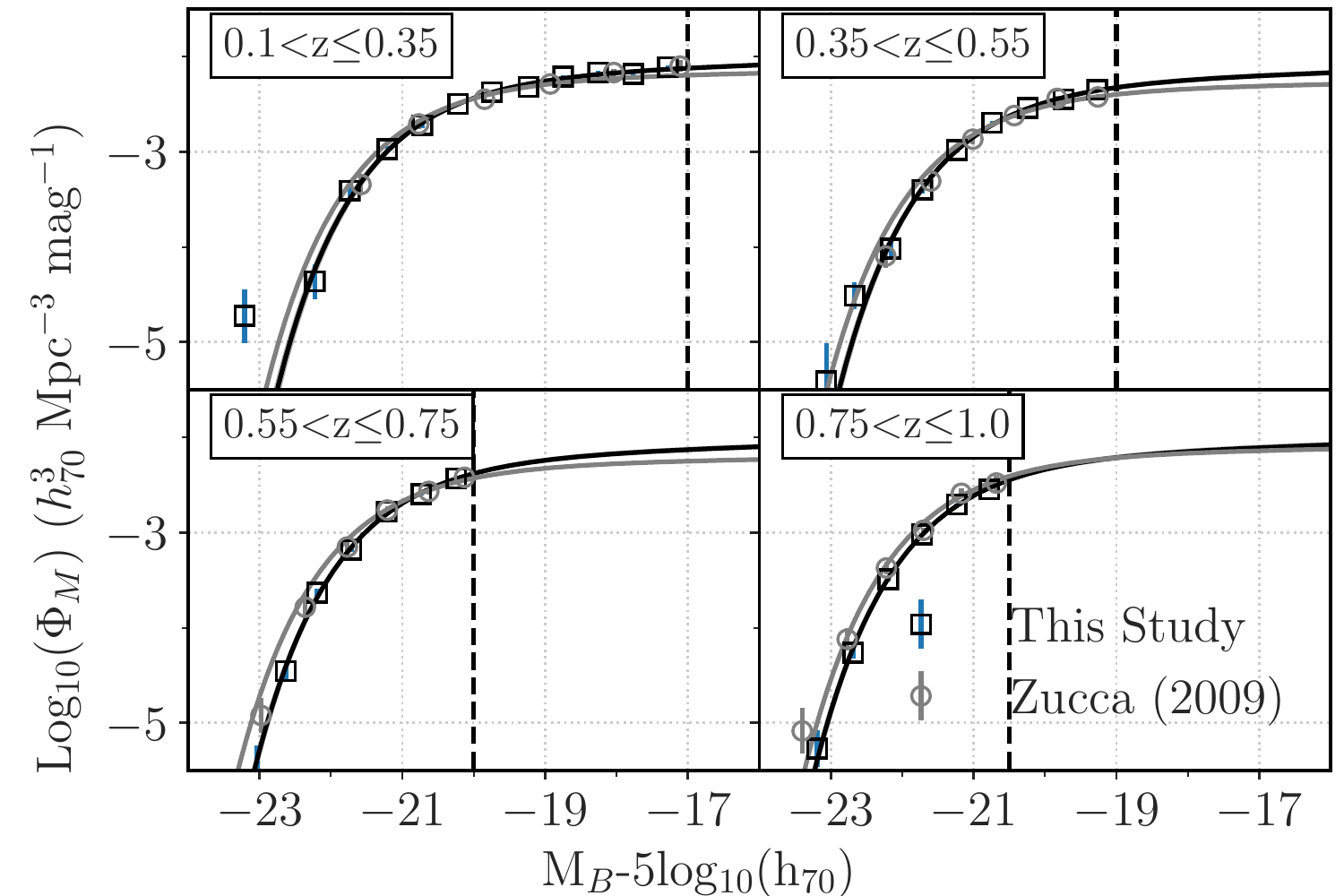}
\caption{Plot comparing our estimate of the total galaxy luminosity function (black) to the estimation from \citet{Zucca2009} (gray). The Schechter fit parameters from our study and \citet{Zucca2009} are listed in Table \ref{lcbg_lf_tab:LCBG_SCHECHTER}. The vertical dashed line is the bias limit from \citet{Zucca2009}}
\label{lcbg_lf_fig:ZUCCALUMFUNC}
\end{figure}
We calculated the luminosity function of all galaxies in the COSMOS field in four bins between $0.1\leq~z\leq1$. We specifically adopt the redshift bins from a previous study on the COSMOS field carried out by \citet{Zucca2009} of $0.1\le~z <0.35$, $0.35\le~z <0.55$, $0.55\le~z <0.75$, and $0.75\le~z <1.0$. We adopt these redshift bins to make a direct comparison and because they contain enough LCBGs to calculate a robust luminosity function in the lowest redshift bin. We also adopt the bias limit \citep{Ilbert2004} cut-off used in \citet{Zucca2009} when calculating the luminosity function in each redshift bin. The bias limit is the absolute magnitude at which a subsample of galaxies would be detectable in a magnitude limited survey, while the rest of the galaxies would be undetected. This bias leads to an underestimate of the faint end slope when calculating the luminosity function using the 1/V$_{max}$ method. Table \ref{tab:bias} lists these bias limits calculated for highest redshift in each redshift bin of this survey. The bias limit in high redshift samples does not allow us to estimate the faint end slope of the luminosity function properly so  we fix the value of $\alpha$ in all redshift bins using the weighted average value in the bins z= $0.1\le~z <0.35$ and $0.35\le~z <0.55$. These values are $\alpha=-1.03$ for all galaxies, and $\alpha=-0.97$ for LCBGs.

\begin{deluxetable}{rr}
\title{Summary of Observations}
\centering
\tablecaption{Luminosity function bias limit estimated from \citet{Zucca2009} \label{tab:bias}}
\tablehead{\colhead{z-bin} & \colhead{Bias Limit}\\
&\colhead{M$_{B}$, mag}}
\startdata
$0.1-0.35$&$-17$\\
$0.35-0.55$&$-19$\\
$0.55-0.75$&$-20$\\
$0.75-1.0$&$-20.5$\\
\enddata
\end{deluxetable}

Figure \ref{lcbg_lf_fig:ZUCCALUMFUNC} shows the points from our 1/V$_{max}$ calculation of the luminsoity function for all galaxies plotted at the average absolute magnitude of their bin, with the error bars on each point indicating the $1 \sigma$ Poisson errors as mentioned in Section \ref{lcbg_lf_sec:LFE}. We used a Markov Chain Monte Carlo (MCMC)  sampler \citep{Foreman-Mackey2013} to fit a Schecter function to the data points with the results shown in Table \ref{lcbg_lf_tab:LCBG_SCHECHTER}.  Figure \ref{lcbg_lf_fig:ZUCCALUMFUNC} also shows the points from \citet{Zucca2009} using the C$^{+}$ method \citep{Zucca1997}, along with the Schechter function estimated by \citet{Zucca2009} using the STY method in gray. The fits in Figure \ref{lcbg_lf_fig:ZUCCALUMFUNC} use a constant value for the faint end slope ($\alpha=-1.03$ for \citet{Zucca2009} and $\alpha=-1.06$ for this study). The curves match within the error.  

We compare the Schechter function parameters fit to our calculation to those from \citet{Zucca2009} in Table \ref{lcbg_lf_tab:LCBG_SCHECHTER}. The $68\%$ confidence values for $\phi^{*}$ and M$_{B}^{*}$ are generally not in agreement, but \citet{Zucca2009} notes that their errors on these parameters are underestimates, and we note that $\phi^{*}$ and M$_{B}^{*}$ are degenerate values that depend on the value selected for $\alpha$. Therefore, we calculate the luminosity density (j$_{B}$) to compare values derived from both studies while avoiding the degeneracy of $\phi^{*}$ and M$_{B}^{*}$. The luminosity density is defined as:
\begin{equation}
j_{B}=\phi^{*}L^{*}(\Gamma(\alpha+2,L_{M_{B}=-18.5})-\Gamma(\alpha+2,L_{M_{B}=-23.5}))
\end{equation}
where $L^{*}$ is the characteristic luminosity calculated from the characteristic magnitude, $\Gamma$ is the upper incomplete gamma function, and L$_{M_{B}=x}$ is the luminosity calculated from the absolute magnitude limits defined above. Calculation of the luminosity density requires integrating the product of the luminosity and the Schechter function. We integrated from M$_{B}=-$23.5 mag, where we have a steep drop off in the number of sources in our catalog, to M$_{B}=-$18.5 mag, the limiting luminosity in the LCBG selection criteria. The luminosity density values between the two fits agree within $2 \sigma$ in the first three redshift bins and only diverge significantly in the last bin. 

\section{The LCBG Luminosity Function}
\label{sec:Frac}

\begin{deluxetable*}{lrrrrrr}
\tabletypesize{\footnotesize}
\tablecaption{Fit parameters to the luminosity function in each redshift bin for all galaxies from \citet{Zucca2009}, all galaxies in this study, and LCBGs in this study \label{lcbg_lf_tab:LCBG_SCHECHTER}}
\tablehead{\colhead{z-bin} & \colhead{Number \tablenotemark{a}} & \colhead{Number\tablenotemark{b}} & \colhead{$\alpha$} &  \colhead{M$_{B}^{*}$-5log($h_{70}$)} & \colhead{$\phi^{*}$} & \colhead{$j_{B}$}\\ 
& & & &\colhead{(mag)} & \colhead{(10$^{-3}~h^{-3}_{70}~$Mpc$^{-3}$)} & \colhead{(10$^{8}~h_{70}$L$_{\odot}$Mpc$^{-3}$)}\\}
\startdata
Zucca\\
$0.1-0.35$&$1968$&$1876$&$-1.03$ (fixed)&$-20.73^{+0.05}_{0.06}$&$6.45^{+0.15}_{-0.15}$&$1.71^{+0.10}_{-0.10}$\\
$0.35-0.55$&$2059$&$1841$&$-1.03$ (fixed)&$-20.91^{+0.05}_{0.05}$&$4.90^{+0.11}_{-0.11}$&$1.57^{+0.08}_{-0.08}$\\
$0.55-0.75$&$2163$&$2086$&$-1.03$ (fixed)&$-21.14^{+0.04}_{0.04}$&$5.57^{+0.12}_{-0.12}$&$2.25^{+0.10}_{-0.10}$\\
$0.75-1.0$&$1769$&$1750$&$-1.03$ (fixed)&$-21.17^{+0.04}_{0.04}$&$7.15^{+0.17}_{-0.17}$&$2.98^{+0.13}_{-0.13}$\\
All Galaxies\\
$0.1-0.35$&$4034$&$3867$&$-1.08^{+0.02}_{-0.03}$&$-20.61^{+0.05}_{-0.06}$&$6.65^{+0.38}_{-0.45}$&$1.51^{+0.05}_{-0.05}$\\
&&&$-1.06$ (fixed)  &$-20.52^{+0.03}_{-0.04}$&$7.30^{+0.14}_{-0.15}$&$1.56^{+0.04}_{-0.04}$\\
$0.35-0.55$&$4143$&$3414$&$-1.04^{+0.06}_{-0.06}$&$-20.70^{+0.07}_{-0.08}$&$5.91^{+47}_{-0.54}$&$1.52^{+0.06}_{-0.06}$\\
&&&$-1.06$ (fixed) &$-20.70^{+0.03}_{-0.03}$&$5.89^{+0.14}_{-0.14}$&$1.52^{+0.03}_{-0.03}$\\
$0.55-0.75$&$3990$&$3341$&$-1.03^{+0.07}_{-0.10}$&$-20.87^{+0.06}_{-0.07}$&$6.86^{+0.46}_{-0.57}$&$2.11^{+0.08}_{-0.08}$\\
&&&$-1.06$ (fixed) &$-20.88^{+0.02}_{-0.03}$&$6.81^{+0.17}_{-0.20}$&$2.10^{+0.03}_{-0.03}$\\
$0.75-1.0$&$2990$&$2863$&$-0.83^{+0.13}_{-0.15}$&$-20.90^{+0.08}_{-0.08}$&$7.74^{+0.49}_{-0.44}$&$2.59^{+0.08}_{-0.03}$\\
&&&$-1.06$ (fixed) &$-21.03^{+0.02}_{-0.02}$&$7.01^{+0.23}_{-0.25}$&$2.53^{+0.04}_{-0.04}$\\
LCBGs \\
$0.1-0.35$&$319$&$319$&$-0.89^{+0.12}_{-0.18}$&$-20.35^{+0.15}_{-0.21}$&$1.62^{+0.26}_{-0.37}$&$0.29^{+0.03}_{-0.02}$\\
&&&$-0.97$ (fixed) &$-20.44^{+0.07}_{-0.09}$&$1.47^{+0.09}_{-0.10}$&$0.30^{+0.02}_{-0.02}$\\
$0.35-0.55$&$880$&$785$&$-1.044^{+0.09}_{-0.13}$&$-20.60^{+0.11}_{-0.15}$&$1.96^{+0.25}_{-0.35}$&$0.46^{+0.03}_{-0.03}$\\
&&&$-0.97$ (fixed) &$-20.53^{+0.05}_{-0.05}$&$2.15^{+0.09}_{-0.10}$&$0.48^{+0.02}_{-0.01}$\\
$0.55-0.75$&$1098$&$935$&$-0.98^{+0.17}_{-0.21}$&$-20.48^{0.10}_{-0.13}$&$3.98^{+0.42}_{-0.44}$&$0.82^{+0.06}_{-0.03}$\\
&&&$-0.97$ (fixed) &$-20.48^{+0.03}_{-0.04}$&$4.00^{+0.18}_{-0.21}$&$0.84^{+0.02}_{-0.02}$\\
$0.75-1.0$&$941$&$894$&$-1.31^{+0.21}_{-0.40}$&$-20.78^{0.14}_{-0.23}$&$7.75^{+0.70}_{-0.96}$&$1.20^{+0.20}_{-0.15}$\\
&&&$-0.97$ (fixed) &$-20.59^{+0.03}_{-0.04}$&$6.92^{+0.31}_{-0.38}$&$1.23^{+0.04}_{-0.04}$\\
\enddata
\tablenotetext{a}{Total number of galaxies in redshift bin}
\tablenotetext{b}{Total number of galaxies brighter than the bias limit}
\end{deluxetable*}

\begin{deluxetable*}{cccccc}
\tablecaption{\label{lcbg_lf_tab:LCBG_LUMINOSITY_FUNCTION}Values for the  LCBG luminosity function in each redshift bin}
\tablehead{\multicolumn{2}{c}{M$_{B}^{*}$-5log($h_{70}$)} & \multicolumn{4}{c}{Luminosity Function Value (10$^{-3}~h^{-3}_{70}~$Mpc$^{-3}$)}\\ 
\colhead{\hspace{0.5cm}Min}\hspace{0.5cm} & \colhead{\hspace{0.5cm}Max}\hspace{0.5cm} & \colhead{\hspace{0.5cm}$0.1 \le z < 0.35$}\hspace{0.5cm} & \colhead{\hspace{0.5cm}$0.35 \le z < 0.55$}\hspace{0.5cm} & \colhead{\hspace{0.5cm}$0.55 \le z < 0.75$}\hspace{0.5cm} & \colhead{\hspace{0.5cm}$0.75 \le z < 1.0$}}
\startdata
$-23.5$ & $-23.0$ & $0.06\pm0.02$ & $...$ & $...$ & $...$\\
$-23.0$ & $-22.5$ & $...$ & $0.01\pm0.01$ & $...$ & $0.01\pm0.0$\\
$-22.5$ & $-22.0$ & $...$ & $0.02\pm0.01$ & $0.03\pm0.01$ & $0.06\pm0.01$\\
$-22.0$ & $-21.5$ & $0.04\pm0.02$ & $0.1\pm0.02$ & $0.19\pm0.02$ & $0.3\pm0.02$\\
$-21.5$ & $-21.0$ & $0.21\pm0.04$ & $0.31\pm0.03$ & $0.53\pm0.03$ & $0.83\pm0.03$\\
$-21.0$ & $-20.5$ & $0.41\pm0.06$ & $0.56\pm0.04$ & $0.99\pm0.04$ & $1.46\pm0.06$\\
$-20.5$ & $-20.0$ & $0.64\pm0.07$ & $0.9\pm0.05$ & $1.66\pm0.06$ & $0.68\pm0.08$\\
$-20.0$ & $-19.5$ & $0.72\pm0.07$ & $1.25\pm0.06$ & $1.34\pm0.08$ & $...$\\
$-19.5$ & $-19.0$ & $0.94\pm0.08$ & $1.38\pm0.07$ & $0.24\pm0.06$ & $...$\\
$-19.0$ & $-18.5$ & $1.04\pm0.09$ & $1.54\pm0.13$ & $...$ & $...$\\
\enddata
\end{deluxetable*}

\begin{figure*}[htb!]
\centering
\includegraphics[width=\textwidth]{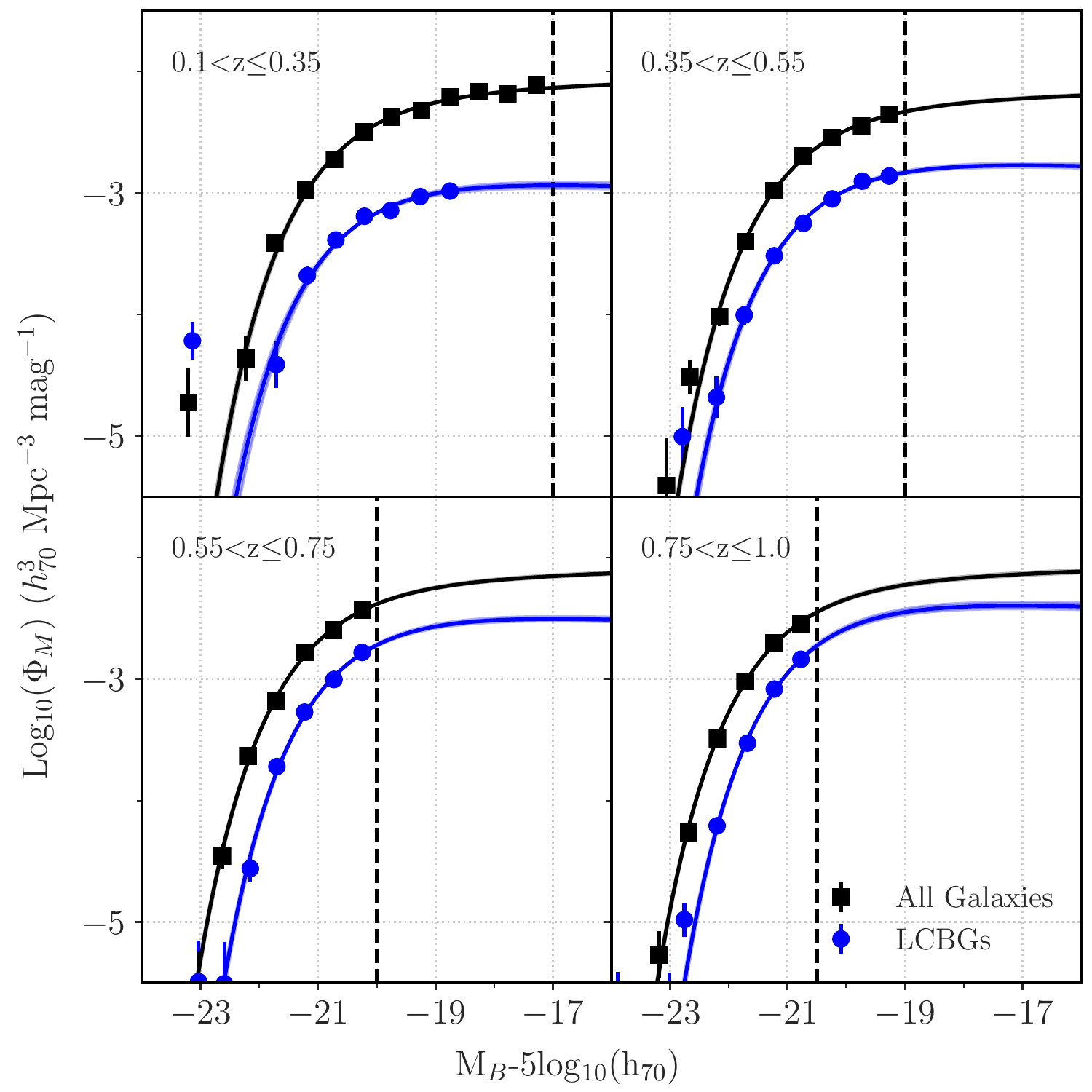}
\caption{Luminosity function for the entire galaxy population (black) and LCBGs (blue). Points mark the average absolute magnitude of sources in that bin, not necessarily the center of the magnitude bin. The vertical dashed line shows the bias limit of the sample, described in Section \ref{lcbg_lf_sec:LFEstimator}}
\label{fig:LCBGLUMFUNC}
\end{figure*}

We show the luminosity function for both the total galaxy population and the LCBG sub-population in Figure \ref{fig:LCBGLUMFUNC}, and their respective Schechter parameters in Table \ref{lcbg_lf_tab:LCBG_SCHECHTER}.  In the LCBG population the characteristic magnitude, M$^{*}$, brightens by roughly $\sim0.2$ mag over the redshift range $0.1\le~z <1.0$. Both the LCBG space density, $\phi^{*}$, and luminosity density, $j_{B}$,  increase by a factor of $\sim4$ over this same redshift range. LCBGs become an important contributor to the total luminosity density, contributing $\sim20\%$ of the B-band luminosity density between $0.1\le~z<0.35 
$ and almost $50\%$ between $0.75\le~z <1.0$. This indicates that the evolution of LCBG number density is not necessarily due to probing different galaxy populations as we do not see much evolution in M$^{*}$, but that there is a large change in number density of the population as previously shown by \citet{Guzman1998}.

\begin{figure}[htb]
\centering
\includegraphics[width=0.5\textwidth]{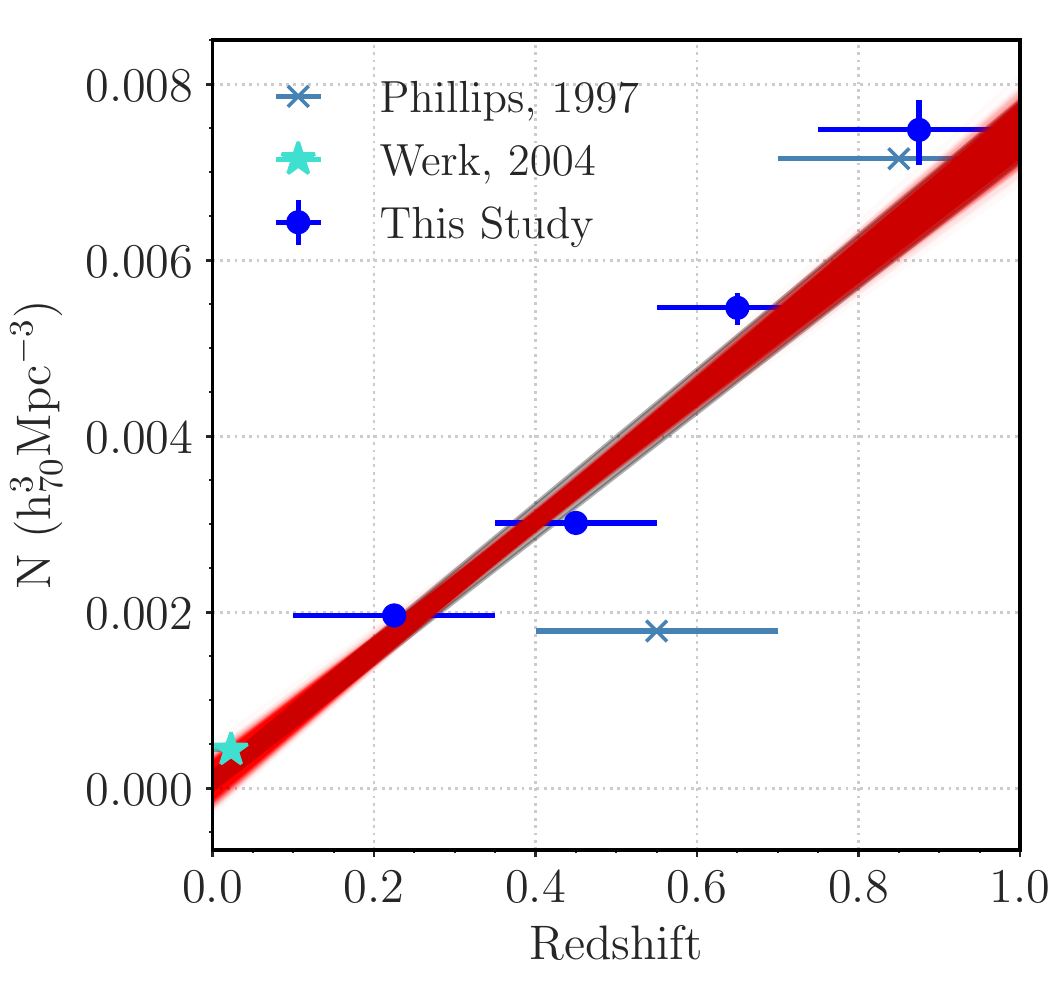}
\caption{Number density of LCBGs in the COSMOS field. We include number density estimates from previous LCBG studies by \citet{Phillips1997}, and \citet{Werk2004}.  We also show the line fit only to data points calculated in this study which is N=$0.007\times~z+0.0001$, with the shaded region showing the 68$\%$ confidence interval of the fit}
\label{fig:NumDensEv}
\end{figure}

We estimate the number density of LCBGs in our four redshift bins by integrating all of the Schechter functions generated in the MCMC analysis from M=$-\infty$ to $-18.5$ mag. We then selected the most likely value as the accepted value and plotted the 68\% confidence interval in the error bars. 

The LCBG number density is plotted against redshift in Figure \ref{fig:NumDensEv}. At the lowest redshift range in our sample, $0.1\le~z <0.35$, the LCBG number density is $2.0\times10^{-3}h_{70}^{3}$Mpc$^{-3}$ and at the highest redshift range, $0.75\le~z <1.0$, the LCBG number density is $7.4\times10^{-3}h_{70}^{3}$Mpc$^{-3}$. Figure \ref{fig:NumDensEv} also includes a linear fit to our number density values using the MCMC approach to better understand the errors in our fit. The fit includes the 68$\%$ confidence interval shaded in red, and approximately 1000 separate fits covering the $68\%$ confidence interval from the MCMC analysis plotted in a darker red shade. The equation, fit to our points is: 
\begin{eqnarray}
N=(7.3\pm0.2  \times z+0.1\pm0.1)\\
\times10^{-3}~h^{3}_{70}~Mpc^{-3}\nonumber
\label{lcbg_lf_eq:numdentrend}
\end{eqnarray}
where N is the number density and z is the redshift. This can be used in future studies to find the estimated LCBG number density at a given redshift.
 
\citet{Werk2004} used the 1/V$_{max}$ method to calculate the number density of LCBGs below z=0.045 and found it to be $4.4\times10^{-4}h_{70}^{3}$Mpc$^{-3}$. This is roughly a factor of 17 lower than the number density at z=1.0 and as seen in Figure \ref{fig:NumDensEv} falls within the 68\% confidence interval of our LCBG number density fit. Figure \ref{fig:NumDensEv} also includes the LCBG number densities estimated by \citet{Phillips1997} and shows that between $0.4\le~z <0.7$ it is about 33\% lower than our value. \citet{Phillips1997} do not include uncertainties in their estimate and suggest a large incompleteness correction to account for missing lower luminosity LCBGs. \citet{Werk2004} suggests that the uncertainty in the number density estimate from \citet{Phillips1997} is at least 33\%, which would make their low redshift value fit closer to our estimate of the number density of LCBGs.

\begin{figure}[htb]
\centering
\includegraphics[width=0.48\textwidth]{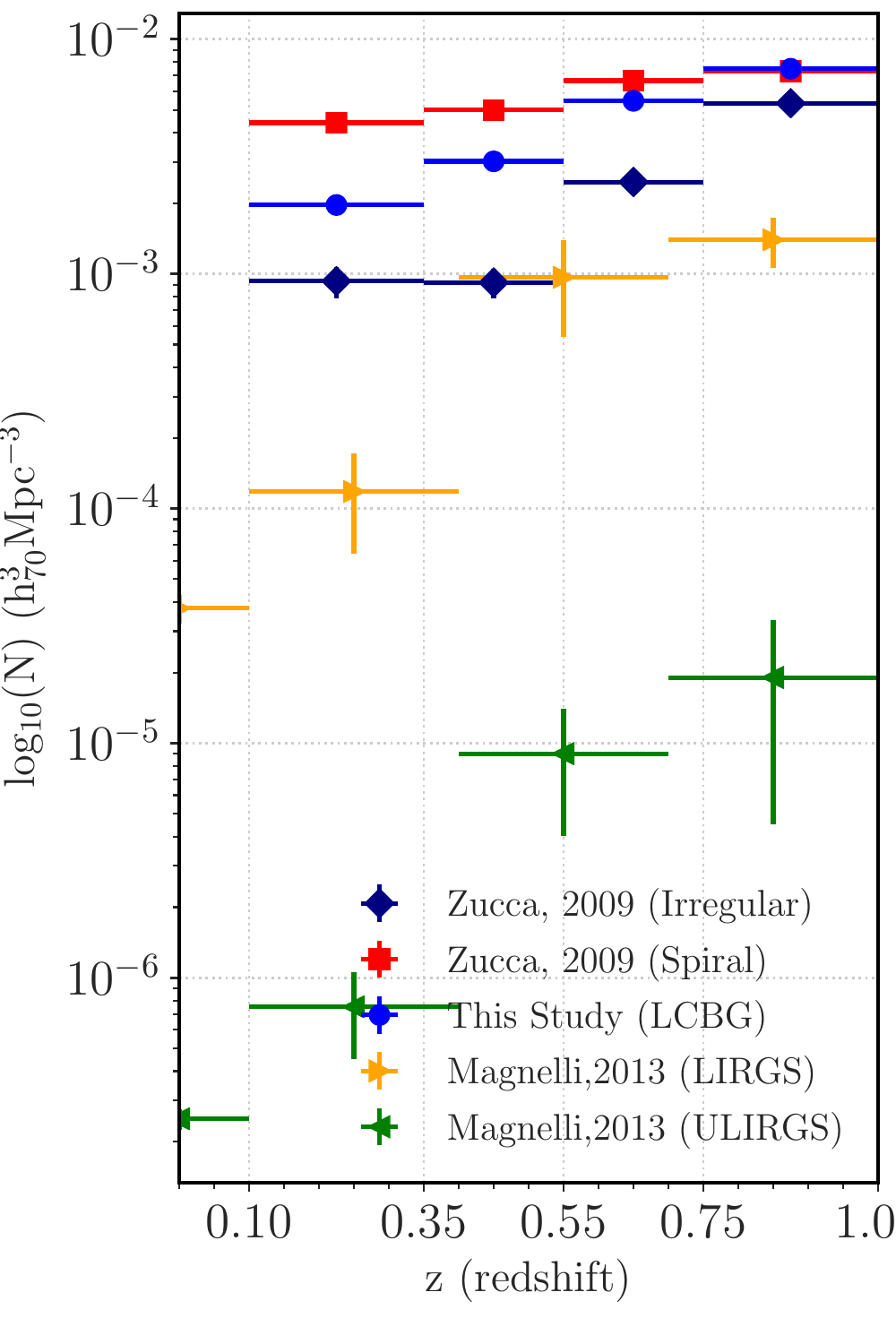}
\caption{Plot comparing the number density evolution of LCBGs to spiral galaxies \citep{Zucca2009}, irregular galxies \citep{Zucca2009}, LIRGs \citep{Magnelli2013} and ULIRGs \citep{Magnelli2013} over the range $0.1\le~z <1.0$}
\label{lcbg_lf_fig:NumDensComp}
\end{figure}

We also compare the evolution of the number density of LCBGs to the number density of other star-forming galaxy populations including luminous infrared galaxies (LIRGs) and ultra luminous infrared galaxies (ULIRGs) found in the Herschel-PACS Far Infrared Survey as calculated in \citet{Magnelli2013}, irregular and spiral galaxies in COSMOS as calculated in \citet{Zucca2009}. Galaxy morphology was determined in the COSMOS field by \citet{Tasca2009}, and \citet{Zucca2009} used that classification to calculate the luminosity function for early type, spiral, and irregular galaxies. We estimate the number density of irregular and spiral galaxies in \citet{Zucca2009} by integrating the luminosity function for the two galaxy morphologies to M$_{B}=-$18.5 mag, the absolute magnitude cut-off for LCBGs. Figure \ref{lcbg_lf_fig:NumDensComp} shows the number density of LCBGs rises more quickly than the number density of spiral galaxies between $0.1\le~z <1.0$ but not quite as rapidly as irregular galaxies. If LCBGs were predominantly spiral type or irregular type galaxies we would expect their number density to change in a way similar to the parent population. The fact that we do not see similar evolution to spiral or irregular galaxies indicates, as previously suggested, that the LCBG population is morphologically heterogeneous and includes both spiral and irregular galaxies. \citet{Magnelli2013} show LIRGs and ULIRGs grow in number density by a factor of $\sim10$ and $\sim25$ respectively between $0.1\le~z <1.0$. Over the same redshift range, LCBGs grow in number density by a factor of $\sim4$, meaning ULIRGs and LIRGs also evolve more rapidly in number density. While LCBGs still appear to be one of the most rapidly evolving galaxy types as suggested by \citet{Rawat2007}, other star-forming galaxy types do seem to evolve more rapidly. \citet{Magnelli2013} also finds that LIRGs and ULIRGs contribute $\sim45\%$ of the total star formation rate density of the universe at z$\sim1$. \citet{Guzman1997} found that LCBGs could contribute up to 45\% of the star formation rate density of the universe at z$\approx0.7$. This indicates that at z$\approx$0.7, LCBGs could contribute as much to the star formation rate density of the universe as LIRGs and ULIRGs while also being a much more significant population of galaxies. 

\begin{figure}[!ht]
\centering
\includegraphics[width=0.48\textwidth]{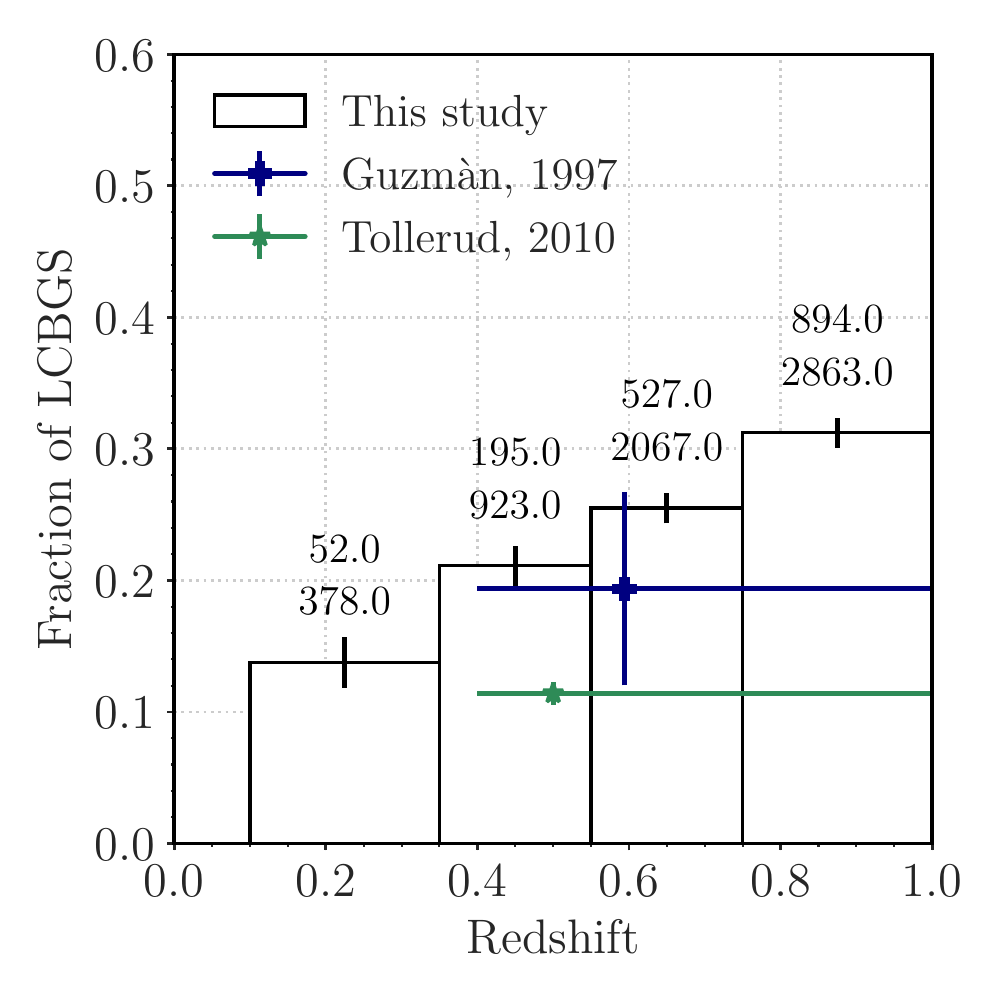}
\caption{The evolution of the fraction of luminous  LCGBs is shown above. Only galaxies more luminous than M$_{B}$<$-$20.5 mag are detected at the redshift limit of this study because of the photometric completeness limit of our dataset. Therefore, in order to compare the fraction of LCBGs in all redshift bins, we calculate the fraction of LCBGs more luminous than M$_{B}$<$-$20.5 mag.  We have also included values from \citet{Guzman1997} and \citet{Tollerud2010} for comparison.}
\label{fig:Fracev_Bright}
\end{figure}

\begin{figure}[!ht]
\centering
\includegraphics[width=0.48\textwidth]{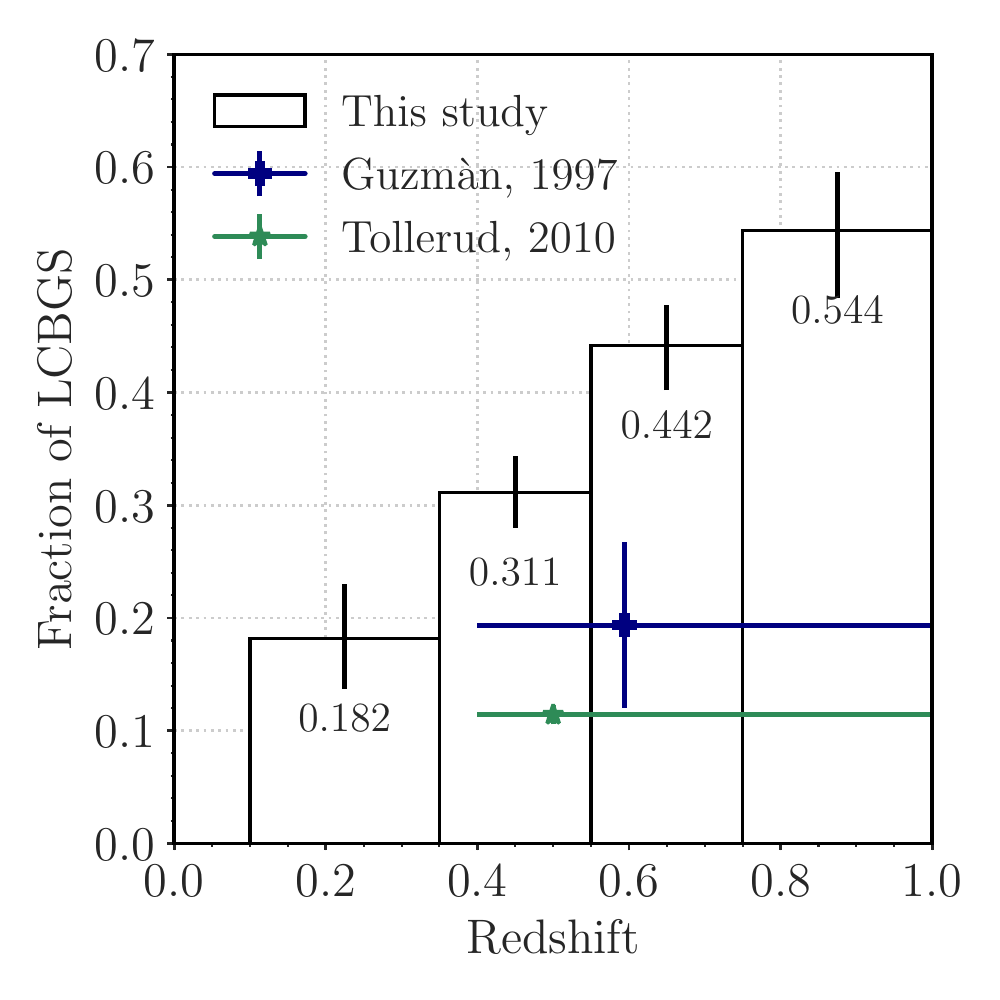}
\caption{An estimation of the evolution of the fraction of all LCBGs. We determined the fraction by integrating the luminosity function of all galaxies to M$_{B}$<$-$18.5 mag, integrating the luminosity function of LCBGs and taking the quotient. We have also included values from \citet{Guzman1997} and \citet{Tollerud2010} for comparison.}
\label{fig:Fracev_All}
\end{figure}

We estimate the fraction of LCBGs in each redshift bin in Figures \ref{fig:Fracev_Bright} and \ref{fig:Fracev_All}. Figure \ref{fig:Fracev_Bright} shows the fraction of galaxies more luminous than M$_{B}=-$20.5 that are LCBGs. This cutoff matches the completeness limit of the survey in the highest redshift bin and allows us to compare the same population of galaxies across all redshift bins. These more luminous LCBGs make up $\sim14\%$ of the luminous galaxy population at z=0.225 and $\sim31\%$  at z=0.875. In figure \ref{fig:Fracev_All} we estimate the fraction of LCBGs in each redshift bin by determining the quotient of the number density of LCBGs and all galaxies more luminous than M$_{B}=-$18.5 by integrating the respective luminosity functions.  We find that LCBGs make up about $18\%$ of galaxies at z=0.225 and $54\%$ of the galaxies at z=0.875. Both estimates show that LCBGs make up a significant portion of the galaxy population at a redshift of $\sim1.0$. 

\citet{Guzman1997} and \citet{Tollerud2010} have both previously looked at the fraction of LCBG like galaxies. \citet{Guzman1997} selected compact star-forming galaxies that have I$_{F814W}\leq22.5$ mag and $\mu_{F814W}\leq22.2$ mag arcsec$^{-2}$ with no selection based on color. Their sample has an average M$_{B}\approx-20.7$ mag and average $\mu_{F814W}\approx20.5$ mag arcsec$^{-2}$ which falls within our determined criteria for LCBG selection. \citet{Tollerud2010} selected LCBGs with M$_{B}\leq-18.5$ mag and B$-$V$\leq0.6$ mag and R$_{e}<3.5$ kpc which at the absolute magnitude limit corresponds to $\mu_{B}\leq22.7$ mag arcsec$^{-2}$. Therefore some of the galaxies contained in these samples would not likely be counted as LCBGs in our sample.

We compare the fraction of luminous LCBGs in COSMOS to the fraction found in \citet{Guzman1997}. As mentioned above, the photometric completeness of \citet{Guzman1997} roughly matched the photometric completeness in COSMOS and we therefore expect a similar fraction of LCBGs in each sample. \citet{Guzman1997} found LCBGs make up (19.3$\pm$7.4)\% of the galaxy population with I$_{F814W}<$22.5 mag between $0.4\le~z <1.0$ with the median redshift of z=0.594. The fraction of LCBGs in G10/COSMOS which has an apparent magnitude limit of Subaru i$^{+}$=22.5 mag between $0.55\le~z <0.75$ is (25$\pm$2)\% which matches the value determined by \citet{Guzman1997} within the errors. 

We have also plotted results from \citet{Tollerud2010} in Figure \ref{fig:Fracev_All}. They estimated the fraction of LCBGs to be (11.4$\pm$0.8)\% also between $0.4\le~z <1.0$ without stating their limiting apparent magnitude. They do note that for their total sample of galaxies the limiting absolute magnitude at z=1 in the r-band is $-18.5$ mag. At z=0.5, their median redshift, their limiting absolute magnitude becomes $-16.5$ mag while the limiting absolute magnitude in the COSMOS field is $-$19 mag. It is therefore not surprising that \citet{Tollerud2010} finds a lower fraction of galaxies that are LCBGs as they are counting fainter galaxies in their total galaxy population than we do.

\section{Conclusion}\label{sec:Conc}
We have traced the evolution of the number density of LCBGs between $0.1\le~z <1.0$ using, for the first time, a consistent homogeneous data set and a sample of LCBGs almost two orders of magnitude larger than any other sample studied. We have done this by deriving the luminosity function of the total galaxy population and of LCBGs in four redshift bins between $0.1\le~z <1.0$. We have found that the LCBG population appears to evolve as rapidly as previously suggested, with the number density at z=0.88 approximately four times larger than that at z=0.18, and approximately 20 times larger than the local number density estimated by \citet{Werk2004}. 

We find M$^{*}$ decreases by $\sim$0.22 mag and $\phi^{*}$  increases by $\sim$5.4$\times$10$^{-3}~h^{-3}_{70}~$Mpc$^{-3}$ for LCBGs in the redshift range $0.1\le~z <1.0$. We also found that the luminosity density increases  by a factor of 4 between $0.1\le~z <1.0$ driven by the change in number density. We also show that LCBGs contribute roughly 48\% of the luminosity density of galaxies more luminous than M$_{B}=-$18.5 at z$\sim$0.88. We find that the LCBG number density evolves as $N=(7.3\pm0.2\times~z+0.1\pm0.1)\times10^{-3} ~h^{-3}_{70}~$Mpc$^{-3}$, in line with previous studies. LCBGs do not evolve as rapidly as irregular galaxies, LIRGs or ULIRGs, but they do evolve more rapidly than spiral galaxies. Finally, we can see that  at z=0.88, LCBGs make up approximately 31\% of the galaxy population more luminous than M$_{B}$=$-$20.5 mag in COSMOS and approximately 54\% of the galaxy population more luminous than M$_{B}=-$18.5 mag. The LCBG fraction is also in line with previous studies and illustrates how ubiquitous these compact star-forming galaxies become at higher redshift.

Acknowledgments:   We thank the anonymous referee for many helpful suggestions which have significantly improved the paper. We would like to thank Dr. Elena Zucca for discussion and information that helped us with deriving the luminosity function. The G10/COSMOS redshift catalogue, photometric catalogue and cutout tool uses data acquired as part of the Cosmic Evolution Survey (COSMOS) project and spectra from observations made with ESO Telescopes at the La Silla or Paranal Observatories under programme ID 175.A-0839. The G10 cutout tool is hosted and maintained by funding from the International Centre for Radio Astronomy Research (ICRAR) at the University of Western Australia. Full details of the data, observation and catalogues can be found in Davies et al. (2015) and Andrews et al. (2016), or on the G10/COSMOS website: cutout.icrar.org/G10/dataRelease.php

MAB acknowledges support from NSF/AST-1517006.  Support for SMC and GW for this work was provided by NASA through grant number AR 15058 from the Space Telescope Science Institute, Based on observations made with the NASA/ESA Hubble Space Telescope, obtained from the data archive at the Space Telescope Science Institute. STScI is operated by the Association of Universities for Research in Astronomy, Inc. under NASA contract NAS 5-26555  DJP and LRH acknowledge partial support from NSF CAREER grant AST-1149491 and AST 1412578.
\software{Astropy \citep{Price-Whelan2018, Robitaille2013},
kcorrect \citep{Blanton2006},
emcee \citep{Foreman-Mackey2013}, matplotlib \citep{Hunter2007}}

\bibliographystyle{aasjournal.bst}
\bibliography{library}

\end{document}